# Bringing the "perfect lens" into focus by near-perfect compensation of losses without gain media

Wyatt Adams[1], Mehdi Sadatgol[1], Xu Zhang[1], and Durdu Ö. Güney[1,2]

[1]Department of Electrical and Computer Engineering, Michigan Technological University, 1400 Townsend Dr., Houghton, MI 49931-1295, USA

[2]Author to whom any correspondence should be addressed.

Email: dguney@mtu.edu

## Abstract

In this paper, the optical properties and imaging performance of a non-ideal Pendry's negative index flat lens with a practical value for loss are studied. Analytical calculations of the optical properties of the lens are performed, and those results are used to further study the lens and corresponding imaging system numerically. The plasmon injection scheme for loss compensation in negative index metamaterials is applied to the results from the imaging system, resulting in a perfect reconstruction of a previously unresolved image that demonstrates sub-diffraction-limit resolution.

## Introduction

Metamaterials provide unprecedented control of light for diverse applications such as wireless communications [1,2], novel optical materials [3-8], optical analog simulators [9,10], photovoltaics [11-13], quantum manipulation of light [14-16], and imaging [17-26], among many others. The extent to which an imaging system is capable of capturing high spatial frequency components of an incoming wave determines its resolution. Those components with spatial frequency greater than $\omega/c$, where $\omega$ is the angular frequency of the wave and $c$ is the speed of light in a medium, constitute evanescent modes that decay rather than propagate. In a conventional imaging system, the image detector is located far enough away from the source so that the evanescent modes are decayed beyond the sensitivity and noise level of the detector, i.e. in the far-field. Consequently, conventional imaging systems can only detect spatial frequencies up to $\omega/c$. This is the so-called diffraction limit first discovered by Abbe [27]. In order to increase the resolution of imaging systems and retain spatial frequency components greater than $\omega/c$, imaging with a slab of negative refractive index material was proposed [17]. This approach relies on the negative index material for focusing of propagating modes and amplification of evanescent modes incident on the slab. Unfortunately, current negative index metamaterial designs are not suitable for optical imaging due to the extreme sensitivity to absorptive losses in the constitutive components [28,29]. A number of metamaterial loss compensation schemes using gain media have been proposed [30-35]. However, the use of gain media for loss compensation can result in instability and spasing [36].

Previously, a loss compensation scheme that provides full compensation in negative index metamaterials without the need for a gain medium was proposed, called the plasmon injection or $\Pi$ scheme [25]. In the $\Pi$ scheme, loss compensation is achieved by coherent excitation of the eigenmodes of a plasmonic negative index metamaterial by superimposing externally injected surface plasmon polaritons (SPPs) with the lossy domestic SPPs in the metamaterial [25,37,38]. Here, in analogy with optical amplifiers, externally injected and domestic SPPs resemble the "pump" and "signal," respectively. Interestingly, in the context of imaging, this purely physical phenomenon for loss compensation in a negative index metamaterial has been shown to be equivalent to a simple spatial filtering algorithm for imaging with a "non-ideal Pendry's negative index flat lens" (referred to as NIFL for short in the rest of the paper) [25].

In the work presented here, a new imaging system centered on a NIFL is proposed which utilizes the $\Pi$ scheme for loss compensation. Unlike the previous near-field negative index flat lens imaging systems, this technique does not benefit from a lossless negative index material and has no gain requirements. The technique developed here is based on a NIFL with a practical value for loss. Recently, a similar spatial filtering approach to countering losses was proposed which also considered tuning the material parameters of the NIFL and surrounding media [39]. However, the optimum values for loss in the NIFL that were assumed are around one to two orders of magnitude lower than what is used to obtain the imaging results presented here. Also, there is little deviation between the optimized material parameters for different spatial frequencies, which suggests the results may be sensitive to any small changes in those values.

## Methods

Fig. 1 shows the block diagram of the $\Pi$ scheme applied to imaging with a NIFL.

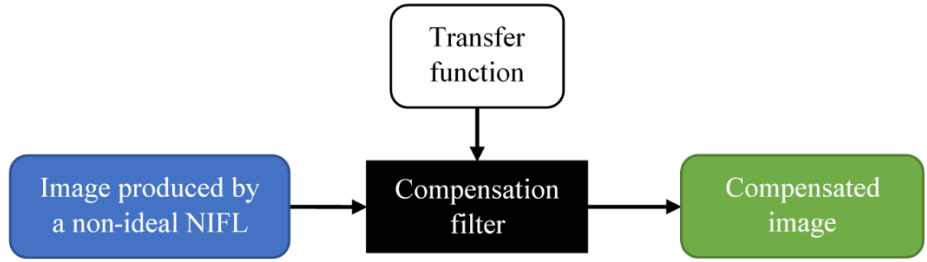

Fig. 1. Block diagram of the Π loss compensation scheme for imaging with a non-ideal Pendry's negative index flat lens.

The procedure begins by producing an image with a NIFL. Then, a filter is applied to the image that compensates the attenuation of the high spatial frequency components. This compensation filter is the inverse of the NIFL transfer function, which can be calculated analytically or numerically. We should note that the method of inverse filtering is well known in the field of image processing, however there are two distinctions to be made between the work presented here and traditional inverse filtering. First, the method in this paper provides compensation for evanescent waves. Secondly, the compensation of these decayed evanescent waves is intimately related to a physical phenomenon for loss compensation in metamaterials as described in [25]. The remainder of this paper explains the methods used to perform the Π scheme loss compensation procedure and form the resulting resolved image.

As previously mentioned, the transfer function of the NIFL can be found through either analytical or numerical calculation. Here, a numerical approach for determining the NIFL transfer function is presented. For any spatial frequency component, the transfer function can be described by the relationship between the electric field at the object plane and image plane. Therefore, in order to find the transfer function it is sufficient to send known plane waves with different spatial frequency $k_y$ and measure the electric field at the image plane. Fig. 2 shows the geometry for the NIFL transfer function calculation using the finite element commercial software package COMSOL Multiphysics. Periodic boundary conditions (PBC) are imposed on the top and bottom boundaries, however the simulation domain itself has limited extent in the y-direction. As a result of the applied PBC in the y-direction, $k_y$ becomes a discretized quantity which can have values of $k_y = \pm m \frac{2\pi}{W_y}$, where $W_y$ is length of the simulation domain in the y-direction and integer $m = 0, 1, 2, ...$ Therefore, an increase in $W_y$ results in more accurate transfer function in terms of number of data points, but also increases the computational domain and in turn the simulation time.

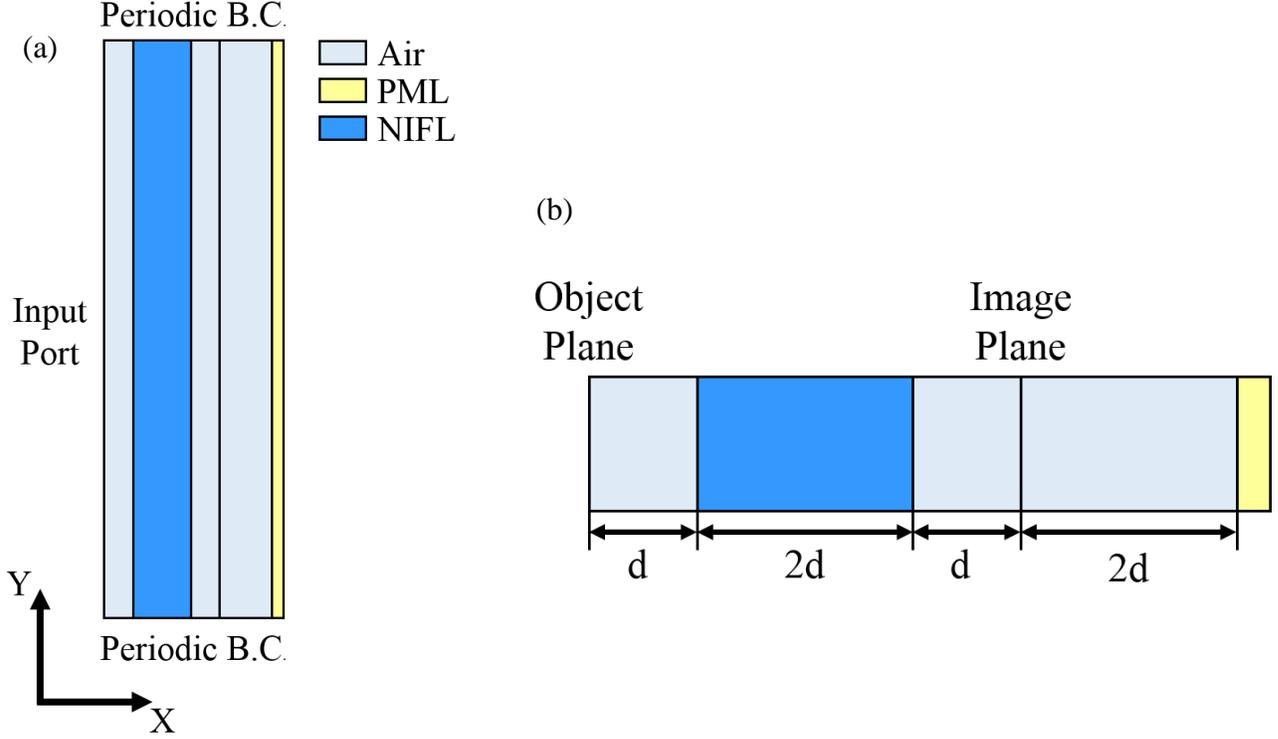

Fig. 2. (a) COMSOL simulation geometry for calculation of the NIFL transfer function. Plane waves with spatial frequency $k_y$ are sent from the input port on the left and measured at the image plane on the opposite side of the NIFL. A perfectly matched layer (PML) is added at the right boundary to suppress the transmitted waves. (b) Cross section of the simulation showing the location of the object and image planes with respect to the NIFL.

After defining the geometric parameters of the NIFL transfer function simulation, the next step is to define the optical properties of the NIFL itself. Consider $\varepsilon_r = -\varepsilon' + j\varepsilon''$ to be the relative complex permittivity and $\mu_r = -\mu' + j\mu''$ to be the relative complex permeability of the NIFL, where $\mu', \mu'', \varepsilon', \varepsilon'' \geq 0$ and $j = \sqrt{-1}$. Then, the refractive index $n$ of the NIFL is

$$n = -\sqrt{\varepsilon_r \mu_r} = -\left(\varepsilon'\mu' - j(\mu'\varepsilon'' + \varepsilon'\mu'') + \varepsilon''\mu''\right)^{1/2}. \qquad (1)$$

Since $\varepsilon'', \mu'' \ll \varepsilon', \mu'$, the $\varepsilon''\mu''$ term can be neglected, and the expression for the refractive index is simplified to

$$n \approx -\left(\varepsilon'\mu' - j(\mu'\varepsilon'' + \varepsilon'\mu'')\right)^{1/2} = -\sqrt{\varepsilon'\mu'}\left(1 - j\left(\frac{\varepsilon''}{\varepsilon'} + \frac{\mu''}{\mu'}\right)\right)^{1/2}. \qquad (2)$$

Using the binomial approximation, Eq. (2) can be further reduced to

$$n \approx \sqrt{\varepsilon'\mu'}\left(-1 + j\frac{1}{2}\left(\frac{\varepsilon''}{\varepsilon'} + \frac{\mu''}{\mu'}\right)\right). \tag{3}$$

If the real part of the relative permittivity and relative permeability are considered to be -1, the relations for the refractive index and impedance $z$ of the NIFL can be written as

$$n = n' + jn'' = -1 + jn'' \approx -1 + j\frac{(\varepsilon'' + \mu'')}{2} \tag{4}$$

and

$$z = \sqrt{\frac{\mu_r}{\varepsilon_r}} = \left(\frac{-1 + j\mu''}{-1 + j\varepsilon''}\right)^{1/2} = \left(\frac{(-1 + j\mu'')(-1 - j\varepsilon'')}{1 + \varepsilon''^2}\right)^{1/2} \approx \frac{1 + j\frac{1}{2}(\varepsilon'' - \mu'')}{1 + \frac{1}{2}\varepsilon''^2} \approx 1 + j\frac{(\varepsilon'' - \mu'')}{2}. \tag{5}$$

Considering the result of Eq. (5), it can be seen that setting $\varepsilon'' = \mu''$ results in an impedance match with free space. However, the effect on the imaging performance of an impedance mismatch introduced when $\varepsilon'' \neq \mu''$ is small compared to the effect of the imaginary part of the refractive index $n''$ in Eq. (4), which characterizes the absorptive loss in the NIFL. As an example to illustrate this, consider the case of $\varepsilon'' = 0.2$ and $\mu'' = 0.1$. From Eqs. (4) and (5), the resulting $n''$ would be 0.15, however the imaginary part of $z$ would be only 0.05. Therefore, for simplicity of analysis the case of $\varepsilon'' = \mu''$ can be chosen without much consideration of the effect of impedance mismatch. By inspection of Fig. 3, it can be determined that ideally the loss in the NIFL would be small in order to preserve the higher spatial frequency components of an image. Unfortunately, fabrication of negative index metamaterials with low loss operating at optical frequencies is difficult. Therefore, an $n''$ of $10^{-1}$ is selected for the rest of the analysis, which is reasonable given current fabricated structures [24]. This corresponds to a figure-of-merit of $|n'/n''| = 10$. Although having such realistic loss levels in the base materials that form the NIFL is sufficient to benefit from the Π scheme, any further improvement in the loss characteristics of the base materials using different techniques [12,30,40-43] can have a profound effect on the Π scheme results.

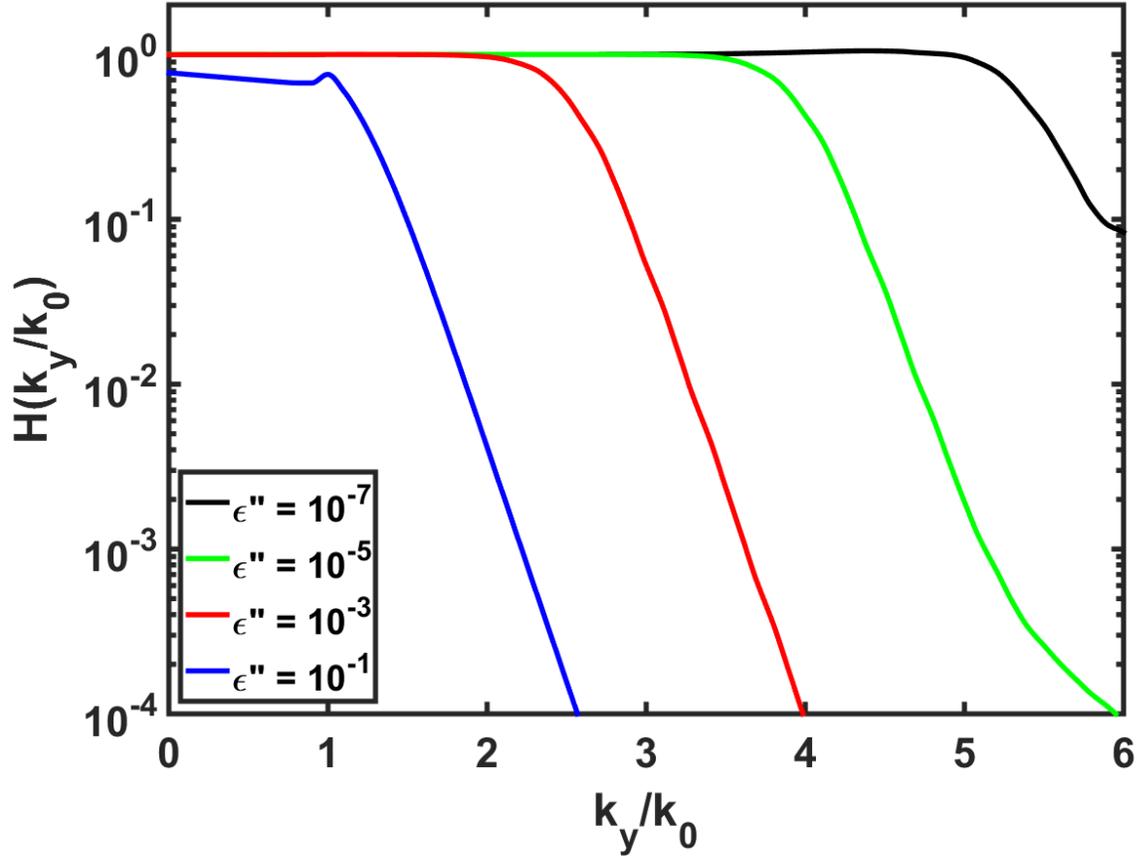

Fig. 3. Transfer function $H(k_y/k_0)$ of the NIFL imaging system for different values of $\varepsilon''$ with $\varepsilon_r = -1 + j\varepsilon''$, $\mu_r = -1 + j\varepsilon''$, $d = 0.25 \mu m$, and $\lambda_0 = 1 \mu m$.

To conclude the methods used here for characterization of the NIFL, a discussion of the effect of the NIFL thickness on the performance of the imaging system is required. Fig. 4 shows the transfer function as the NIFL thickness $2d$ changes from $\lambda_0/2$ to $2\lambda_0$. As the results suggest, a decrease in the NIFL thickness reduces the attenuation of high spatial frequency components, which in turn increases the resolution of the imaging system. However, the thickness of the NIFL cannot be decreased arbitrarily for two reasons. First, the NIFL would be constituted by a metamaterial structure, the minimum thickness of which would be constrained by the size of the corresponding unit cell. Secondly, as the thickness of the NIFL decreases, the working distance of the lens also decreases, making mechanical alignment of the imaging system more difficult.

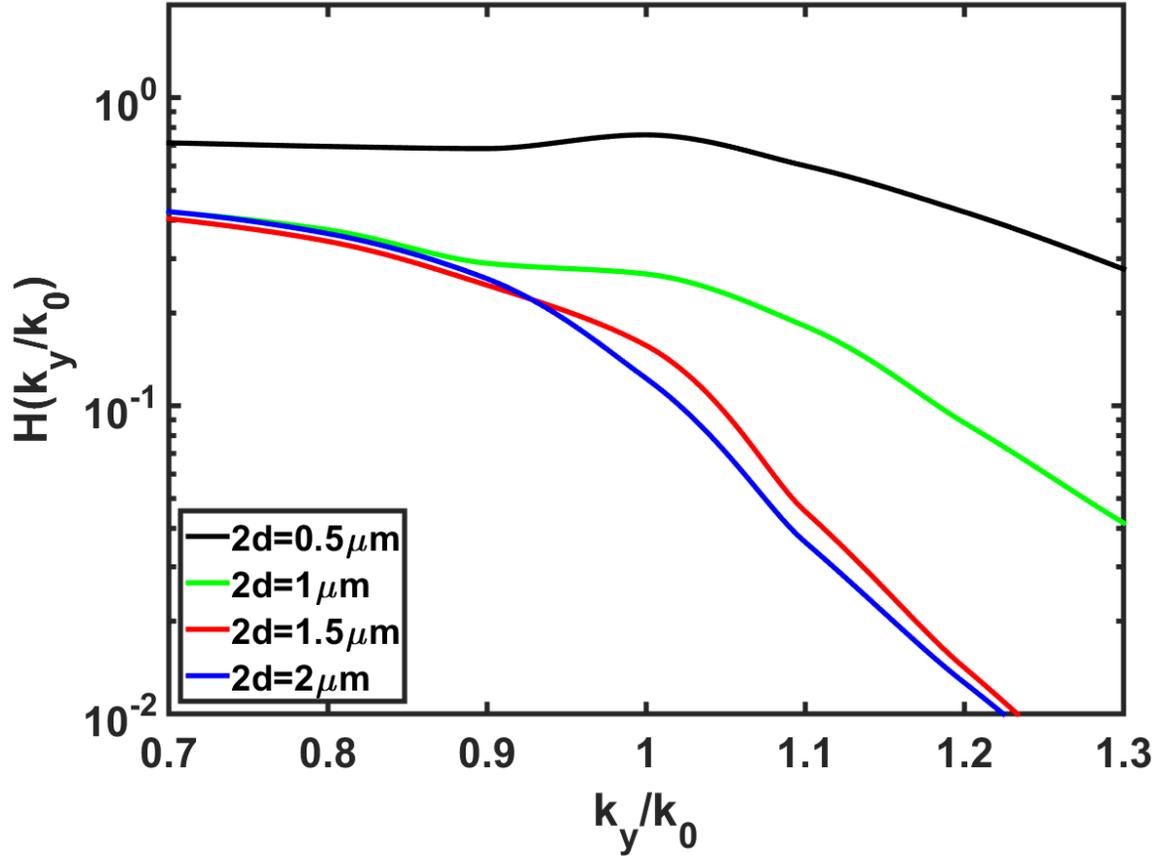

Fig. 4. Change in the transfer function around $k_y/k_0 = 1$ ($k_0 = \frac{2\pi}{\lambda_0}$) as the NIFL thickness $2d$ is changed from $\lambda_0/2 = 0.5 \mu m$ to $2\lambda_0 = 2 \mu m$. The results suggest the employment of a thinner NIFL will result in better imaging performance.

After characterizing the NIFL itself, the next step is to numerically evaluate the imaging performance. Figs. 5 (a) and (b) show the simulation geometry and material settings used to produce an image of some arbitrary object with the NIFL using COMSOL Multiphysics. The object is formed by defining the z-component of the electric field $E_z$ over the object plane, and image is produced by recording $E_z$ on the image plane. In Fig. 5 (c), an object with three Gaussian features separated by $1\mu m$ is defined on the object plane, and the corresponding electric field on the image plane is recorded. Fig. 5 (d) shows a surface plot of the resulting field distribution over the simulation domain.

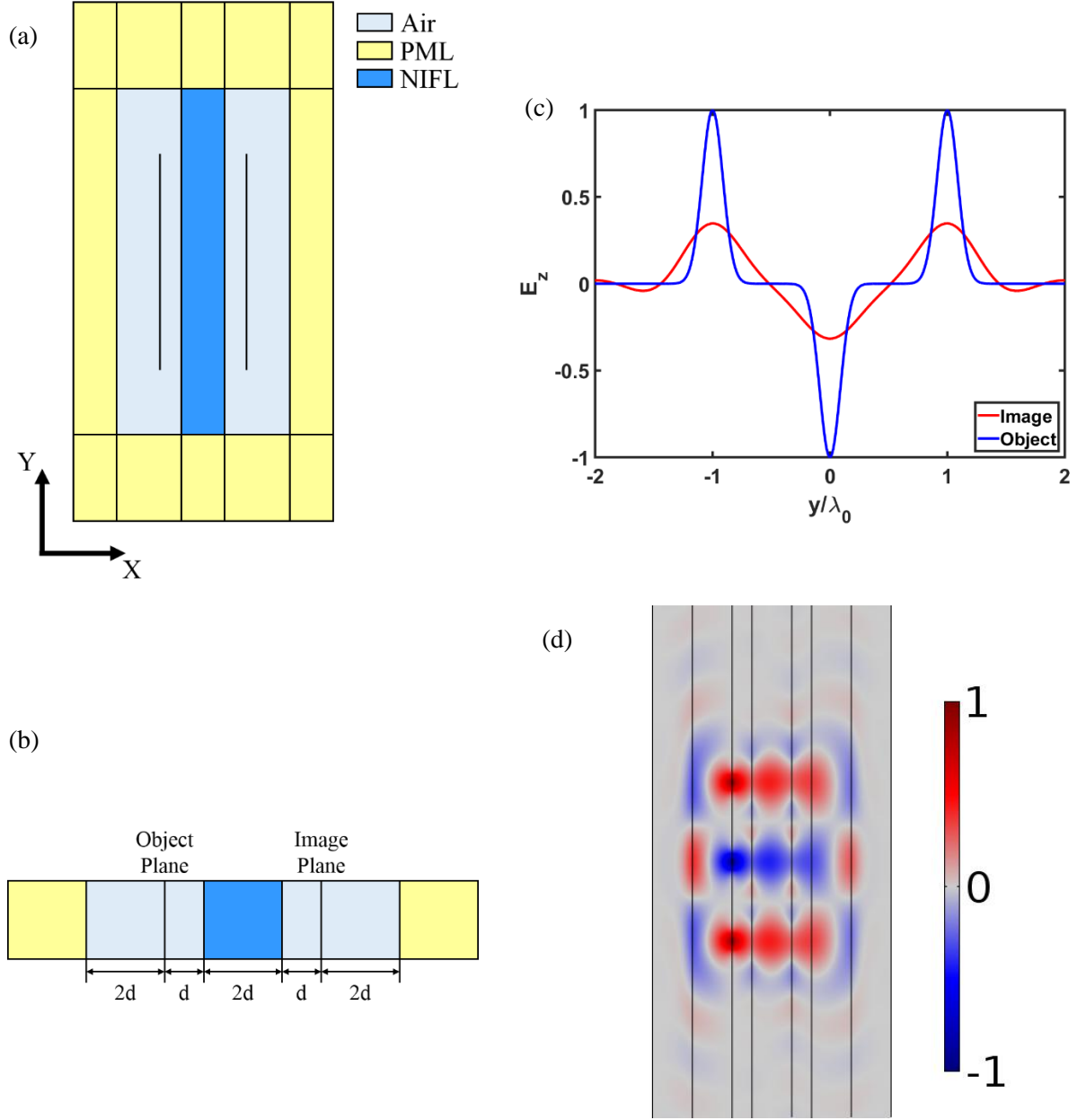

Fig. 5. (a) The geometry and material settings used to perform numerical simulation of the NIFL imaging system. The extent of the model in the y-direction is chosen to be $24\mu m$, though the figure shown here is compressed in the y-dimension to better fit the page. (b) A cross section of the imaging system showing the working distance of the object and image planes from the NIFL. (c) The z-component of electric field $E_z$ on the object and image planes with incident wavelength $\lambda_0 = 1\mu m$ and $2d = 0.5\mu m$. (d) Surface plot of the $E_z$ distribution over the imaging system simulation domain.

This imaging simulation can be repeated to produce the image from any object with arbitrary feature size. Once the image is formed, the resolution can be improved by applying the Π scheme for compensation of losses in the NIFL.

## Results

In order to improve the resolution of the image obtained by the NIFL, it is important to amplify the suppressed spatial frequency components. A compensation filter is required to undo this attenuation made by the imaging system. Obviously, a proper choice for the compensation filter would be the inverse of the imaging system transfer function. This corresponds to the Π scheme loss compensation technique for imaging, where a portion (i.e., pump or auxiliary object) of the total incident field in the object plane can be thought of as coherently exciting the underlying modes of the system in order to compensate the losses in the other portion (i.e., signal or actual object to be imaged) [25]. The equivalent is applying a filter in the spatial frequency domain that amplifies the components with $k_y > k_0$. Fig. 6 (a) shows the compensation filter for the NIFL imaging system described in Fig. 5. As an example, an object with features separated by a distance $\lambda_0/4$, twice beyond the diffraction limit, was imaged by the NIFL. The results of this procedure are shown in Fig. 6 (b). It can be seen that the sub-diffraction-limit features of the object are not resolved in the raw image produced by the NIFL. However, after applying the compensation filter a perfect reconstruction of the original object is achieved.

This procedure can be replicated for any arbitrary object field, provided that enough of the spatial frequency components required to reproduce the field are available to be compensated by the post-processing. Therefore, the limitation to the smallest feature size one could resolve with this technique would solely be the noise floor of the detection mechanism at the image plane, in this case the numerical simulation. Since inverse filtering is prone to noise amplification, it is required to roll off or truncate the filter at some spatial frequency where the noise floor is reached on the image plane. In Fig. 6 (a), it can be seen that the raw image spectrum begins to flatten around $k_y = 2.5k_0$ to $3k_0$. Therefore, simply truncating the filter at $3k_0$ gives a good compensated image that avoids noise amplification at high-$k_y$. It is important to note that truncating the filter in this way requires no *a priori* knowledge of the object; only the detected raw image is needed. While this noise limitation is present in practice, there is no theoretical limit imposed on the compensation scheme presented here.

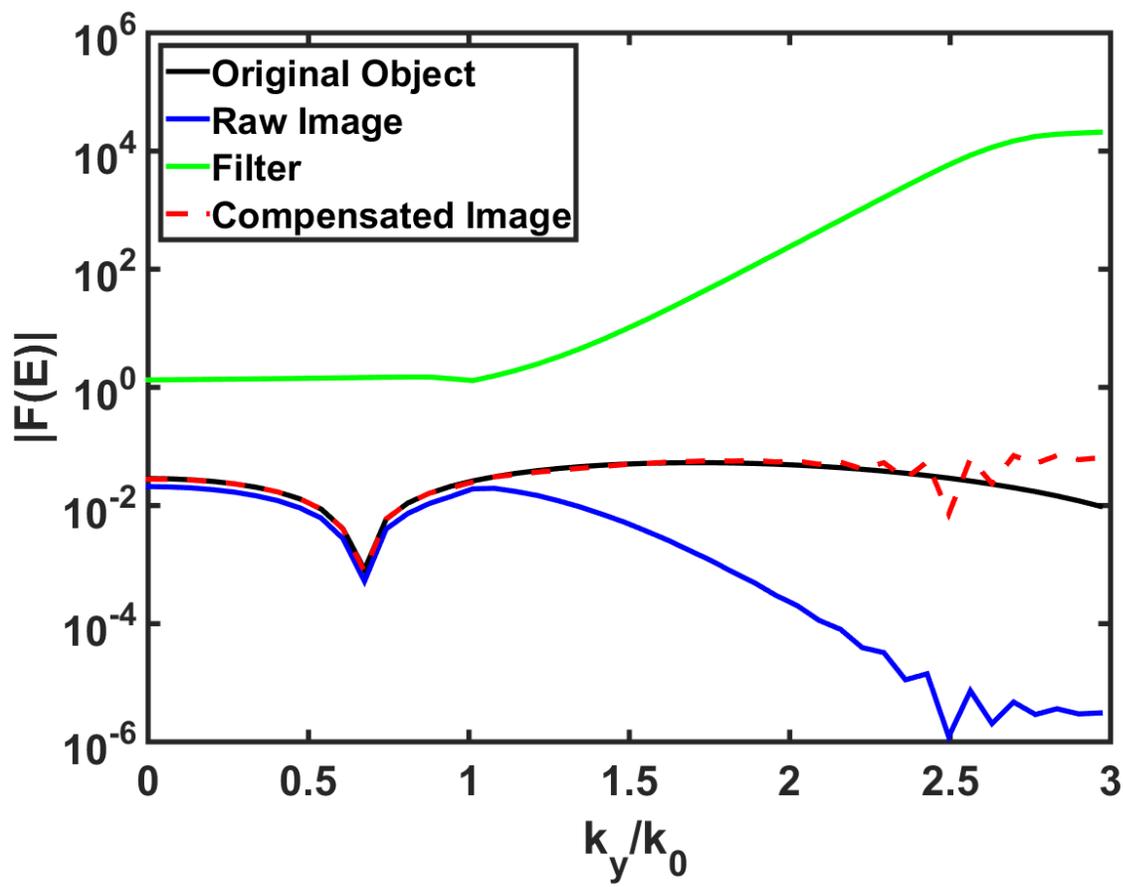

(a)

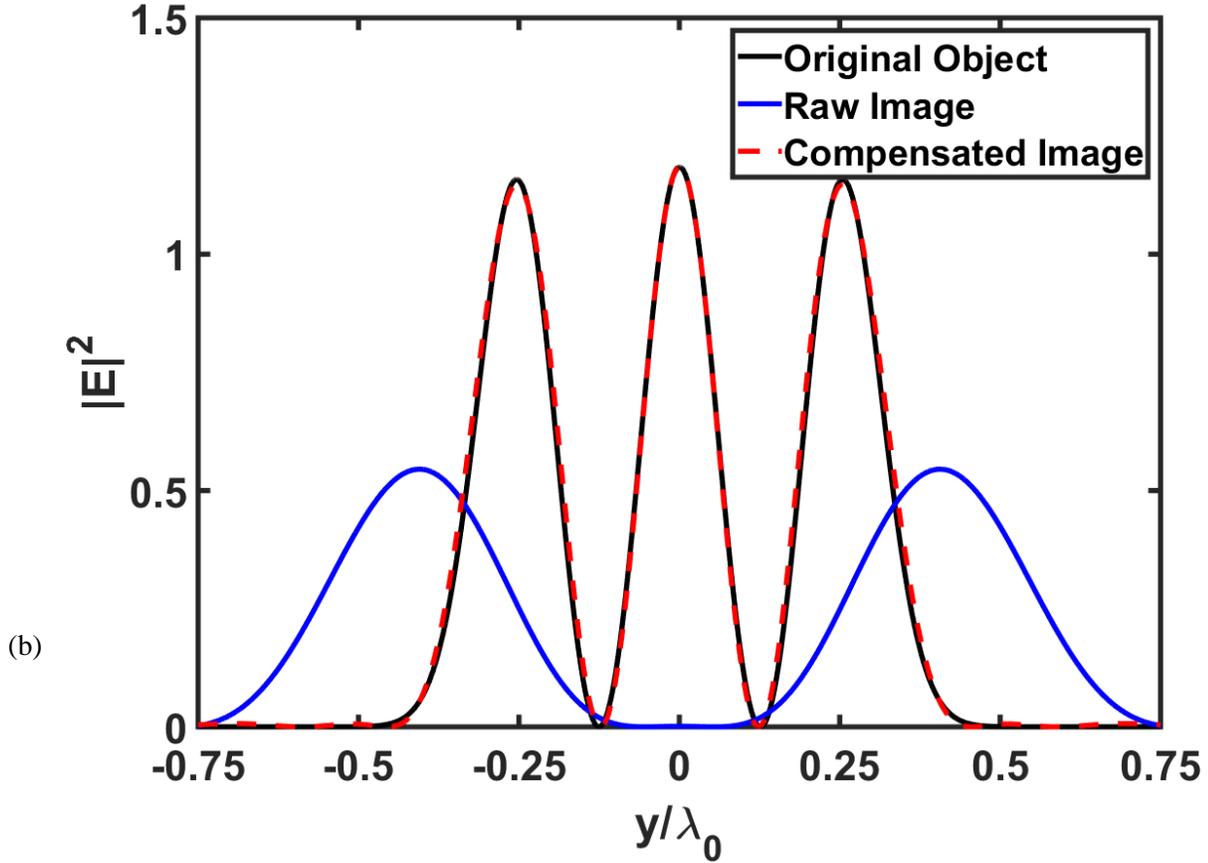

Fig. 6. (a) Fourier spectra for the object, raw image produced by the NIFL, and compensated image with the corresponding filter $H\left(k_y/k_0\right)^{-1}$ resulting from an object with three Gaussian features separated by a distance $\lambda_0/4$. (b) Electric field intensities for the original object, raw image produced by the NIFL, and the compensated image after applying the filter shown in (a).

### Conclusion

In this paper, a study of the optical characteristics and near-field imaging performance of a NIFL with a practical value for loss was performed. The optical properties of the NIFL were investigated analytically, and a numerical calculation of the transfer function was performed and studied. The simulation results yielded an unresolved image from the NIFL, which subsequently underwent loss compensation using the Π scheme from [25]. Applying the Π scheme to a NIFL imaging system involves the simple post-processing step of multiplying the raw image produced by the NIFL with the inverse of the transfer function in the Fourier domain. There are no requirements for electric or magnetic gain in the NIFL and surrounding media. The demonstrated result is a perfect reconstructed image with sub-diffraction-limit feature size. Our findings decouple the more-than-a-decade-long loss problem from the general problem of how to realize a practical "perfect lens" operating in the optical frequencies, and reduce the problem mainly to amenable design and fabrication issues [44-51]. Further developments in metamaterials and the Π scheme approach can lead to advances in other applications besides ultra-high resolution imaging such as photolithography and optical storage technologies.

## Acknowledgments

This work was partially supported by Office of Naval Research (award N00014-15-1-2684) and by the National Science Foundation (NSF) under Grant No. ECCS-1202443. M. S. gratefully acknowledges support from NSF.